\documentclass[twocolumn,showpacs,preprintnumbers,amsmath,amssymb,aps,prb,longbibliography,superscriptaddress]{revtex4-2}

\pdfoutput=1
\usepackage[pdftex]{graphicx}

\usepackage[english]{babel}
\usepackage{hyperref}
\usepackage{amsmath}
\usepackage{float}
\usepackage{soul}

\usepackage{amssymb}
\usepackage[utf8]{inputenc}
\usepackage[T1]{fontenc}
\usepackage{graphicx}
\usepackage{braket}

\usepackage[caption=false]{subfig}

\usepackage{cancel}
\usepackage{color,soul}

\usepackage{epstopdf}
\usepackage{enumitem,kantlipsum}
\usepackage{physics}
\usepackage{bbold}
\usepackage{array}

\begin{document}

\title{Matryoshka approach to Sine-Cosine topological models}

\author{R. G. Dias}
\affiliation{Physics Department $\&$ i3N, University of Aveiro, 3810-193 Aveiro, Portugal}

\author{A. M. Marques}
\affiliation{Department of Physics $\&$ i3N, University of Aveiro, 3810-193 Aveiro, Portugal}

\date{\today}

\begin{abstract}
We address a particular set of extended Su-Schrieffer-Heeger models with $2n$ sites in the unit cell [SSH($2n$)], that we designate by Sine-Cosine models [SC$(n)$], with hopping terms  defined as a sequence of $n$ sine-cosine pairs of the form $\{\sin(\theta_j),\cos(\theta_j)\}$, $j=1, \cdots,n$. These models, when squared, generate a block-diagonal matrix representation  with one of the blocks corresponding to a chain with uniform local potentials. We further focus our study on the subset of SC$(2^{n-1})$ chains that, when squared an arbitrary number of times (up to $n$), always generate a  block  which is again a Sine-Cosine model, if an energy shift is applied and if the energy unit is renormalized. We show that these $n$-times squarable models [SSC$(n)$] and their band structure are   uniquely determined by the sequence of energy unit renormalizations 
and by the energy shifts associated to each step of the squaring process. Chiral symmetry is present in all  Sine-Cosine chains and edge states levels at the respective central gaps are protected by it.  Zero-energy edge states in a SSC$(j)$ chain (with $j<n$) of the Matryoshka sequence obtained squaring the SSC$(n)$ chain with open boundary conditions (OBC), become finite energy edge states in non-central band gaps of the SSC$(n)$ chain. The extension to higher dimensions is discussed.
\end{abstract}

\maketitle

The characterization of  square-root topological insulators ($\sqrt{\text{TI}}$)  relies on the fact that the square of the matrix representation of a $\sqrt{\text{TI}}$  Hamiltonian in the Wannier basis is a block diagonal matrix, more precisely, it is the direct sum $H^2=H_\text{TI} \oplus H_2$ of two blocks $H_\text{TI}$ and $H_2$ that have the same finite energy spectrum, after applying a constant energy downshift, but different eigenstates ($H_\text{TI}$  being the Hamiltonian of a known topological insulator) \cite{Arkinstall2017,Pelegri2019a, Kremer2020}. This reflects the fact that the  $\sqrt{\text{TI}}$ Hamiltonian H is defined in a bipartite lattice [lattice with sublattices A and B, such that the Hamiltonian can be written as a sum of hopping terms (which imply finite Hamiltonian matrix elements) between different sublattices, $H= H_{AB} + H_{BA}$]. 

As very recent examples of squared-root topological insulators, one may cite the diamond chain in the presence of magnetic flux \cite{Kremer2020} or our work on the $t_1t_1t_2t_2$ tight-binding chain (where a modified Zak’s phase, a sublattice chiral-like symmetry, modified polarization quantization, etc., were found \cite{Marques2019}) where $H_{TI}$ corresponds to the well-known  Su-Schrieffer-Heeger (SSH) model \cite{Asboth2016}. In these cases, the topological invariants and symmetries of the SSH Hamiltonian $H_{TI}$ map into modified topological invariants of the original Hamiltonian (see Ref. \cite{Marques2019}). The $t_1t_1t_2t_2$ tight-binding chain is a particular case of the SSH(4) model \cite{Marques2020}  which is a generalization of the topological SSH chain \cite{Eliashvili2017,Maffei2018,Xie2019}. 

Recently, several methods of generating the square root Hamiltonian of a given Topological insulator Hamiltonian in 1D \cite{Arkinstall2017,Kremer2020,Pelegri2019a,Ezawa2020,Ke2020} and 2D \cite{Song2020,Mizoguchi2020,Yan2020,Mizoguchi2021} have been proposed. These methods do not allow its consecutive application due to the appearance of non-uniform local potentials and the consequent loss of the bipartite property. This also reflects the fact that the square-root lattice and the original one are not self-similar.

In this paper, we consider  a particular subset of 1D SSH($N$) models, $N$ being the number of sites in the unit cell,  that we designate by Sine-Cosine models, such that the consecutive squaring of the Hamiltonian has always a block-diagonal matrix representation  with one of the blocks corresponding to a bipartite chain (apart from an energy shift) which is self-similar to the original chain, that is, it is again a sine-cosine model provided that the energy unit is renormalized. The sequence of energy unit renormalizations associated to each step of the squaring process determines the energy gaps in the spectrum of the original chain. Finite energy edge states may be generated at each of these gaps in the case of chains with open boundary conditions and these edge states are protected by a sequence of $n$-chiral symmetries that become evident when $n$-times squaring, energy shifting and renormalizing the Hamiltonian.

The higher dimension generalizations of these 1D models will also have the energy gaps at  the inversion-invariant momenta determined by the renormalization factors.
We show that a square-root Hamiltonian of these higher dimensional models can be also obtained from the 1D counterparts introducing a $\pi$-flux per plaquette.

\begin{figure}[t]
	\centering
	\includegraphics[width=0.90\linewidth]{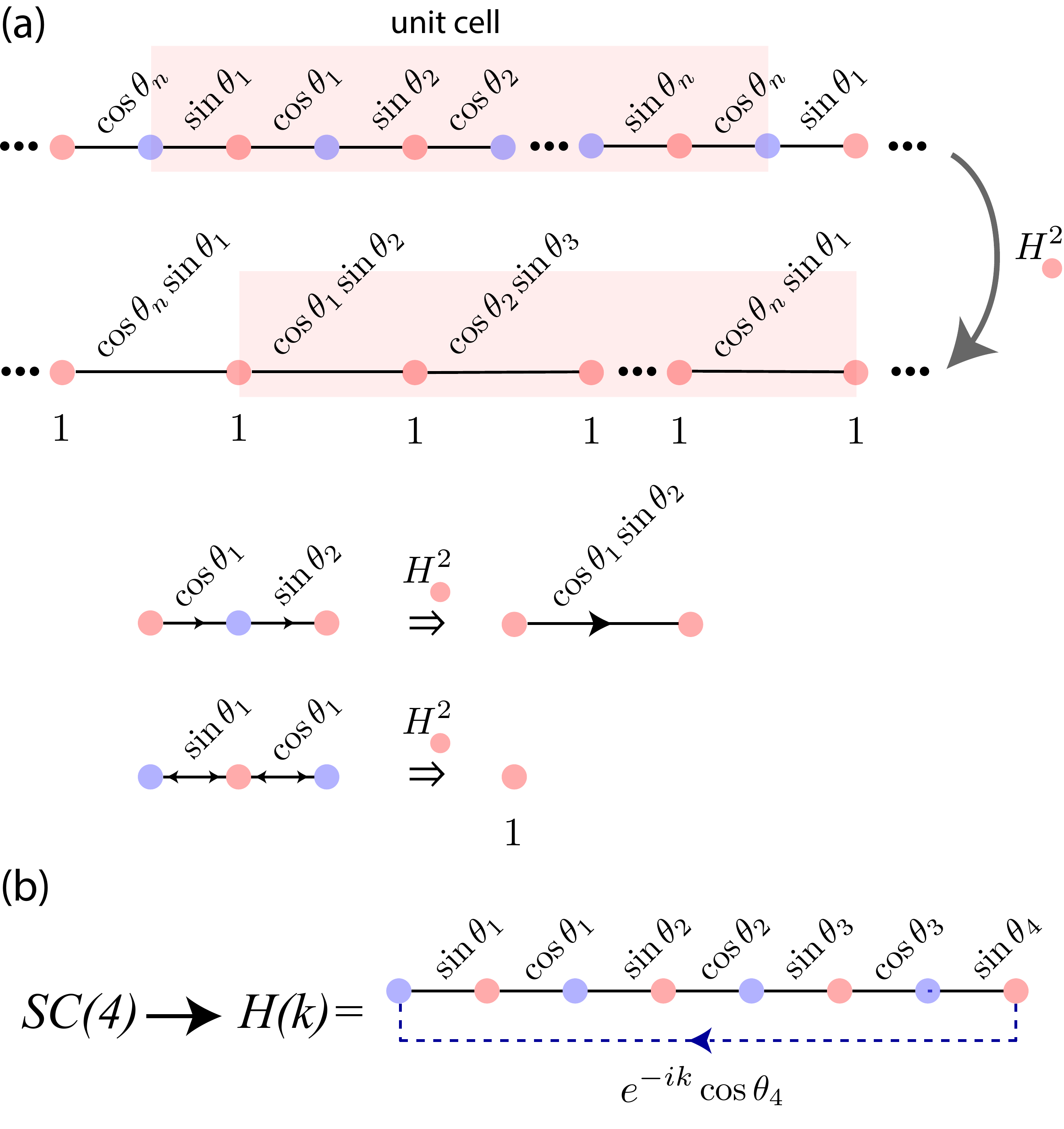}
	\caption{(a) The top diagram illustrates the Sine-Cosine chain with a unit cell of  $2n$ sites and hopping terms  $t_{2j-1}=\sin(\theta_j)$ and $t_{2j}=\cos(\theta_j)$, with $j=1, \cdots,n$. Upon squaring, the Hamiltonian  for one of the sublattices has the form shown in the bottom diagram, with uniform local potentials $\varepsilon_j =\sin(\theta_j)^2+\cos(\theta_j)^2=1$, as illustrated in the simple case of three-site chains in the bottom diagrams of (a) (b) Schematic representation of a SC$(4)$  bulk Hamiltonian. The sites correspondent to A and B sublattices are colored in light blue and light red, respectively.}
	\label{fig:sinecosine}
\end{figure}

\textbf{\emph{Sine-Cosine chains}: }
Assume an SSH($2n$) chain  with a unit cell with  $2n$ sites and with nearest-neighbor hopping terms $t_i$, $i=1, \cdots,$ $2n$,  for some positive integer $n$. The Sine-Cosine model of order $n$, SC$(n)$, is defined imposing that $t_{2j-1}=\sin(\theta_j)$ and $t_{2j}=\cos(\theta_j)$, with $j=1, \cdots,$ $n$ (see top diagram in Fig.~\ref{fig:sinecosine}).

By squaring this bipartite Hamiltonian, one obtains a block-diagonal matrix (one for each sublattice of the bipartite chain) and one of the blocks   [shown in the middle diagram of Fig.~\ref{fig:sinecosine}(a)] corresponds to a tight-binding model with uniform local potentials $\varepsilon_j =\sin(\theta_j)^2+\cos(\theta_j)^2=1$ and hopping terms $t_j=\cos(\theta_j)\sin(\theta_{j+1})$. The uniform potentials can be removed applying  an energy shift of one. Note that if the hopping terms are globally multiplied by a hopping factor $t$, one can still recover the Sine-Cosine form for the Hamiltonian setting this parameter $t$ as the unit energy so that the energy shift is again one (in units of $t$).

\begin{figure}[tb]
	\centering
	\includegraphics[width=0.90\linewidth]{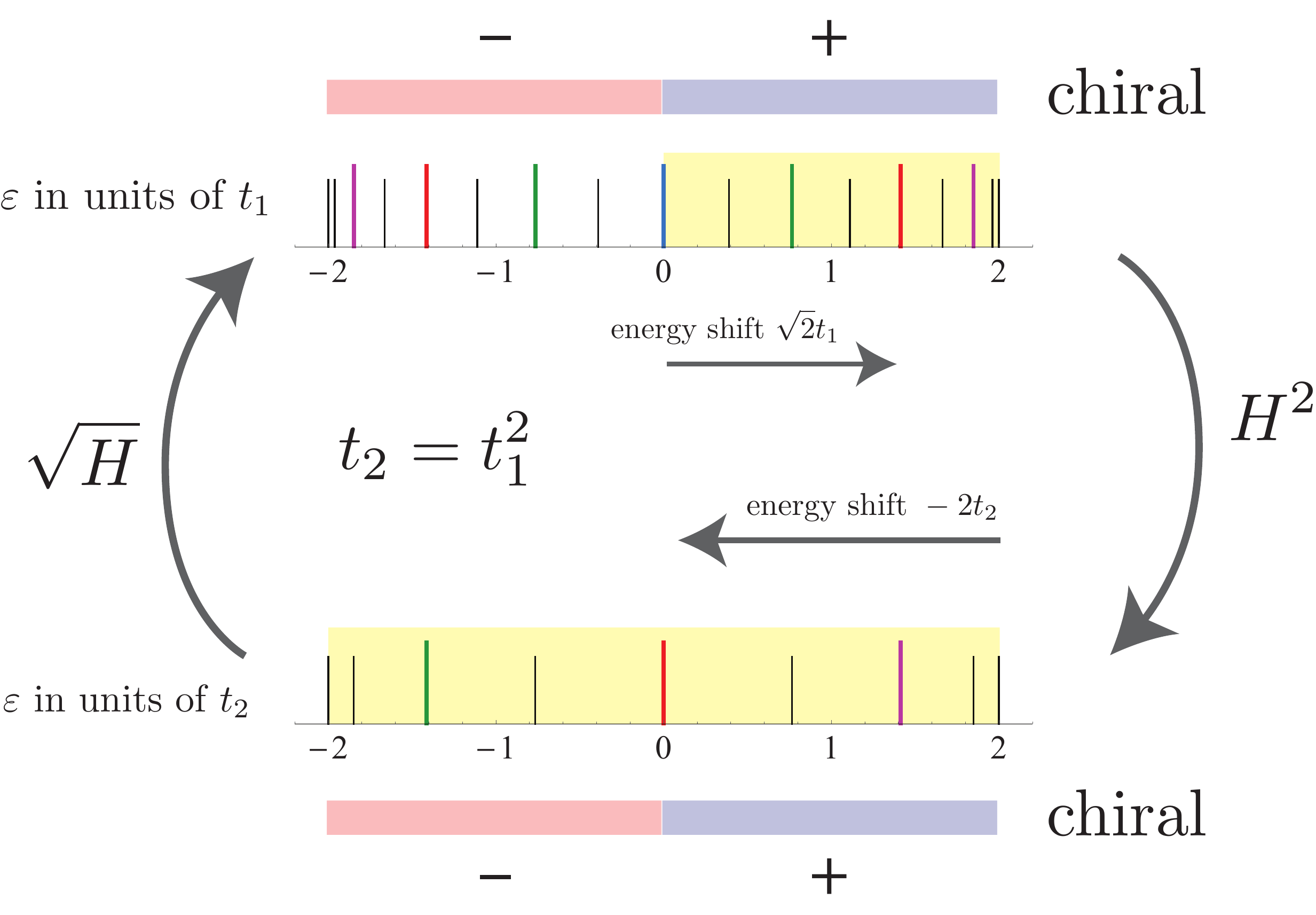}
	\caption{The squaring process for a uniform chain (that is, $\theta_j=\pi/4$, for all $j$)  with PBC. The top spectrum is for a chain with 32 sites and hopping parameter $t_1$  and the bottom for a chain with 16 sites and hopping parameter $t_2$. The colored lines indicate the folding energies in the following steps of the squaring sequence.}
	\label{fig:uniform}
\end{figure}

In the simple case of a uniform chain with hopping parameter $t_1$, one has $\theta_j=\pi/4$, for all $j$, and the hopping parameter is $t_1=t/\sqrt{2}$, so the energy shift necessary remove the uniform potentials (which is one in units of $t$) becomes $2t_1^2$ (see Fig.~\ref{fig:uniform}). Obviously, the sublattice Hamiltonian corresponds to  another uniform chain with $t_2=t^2 \cos(\pi/4)\sin(\pi/4)=t^2/2=t_1^2$.
Note that if we consider the respective inverse operation, the square-root of the $t_2$ chain, the bottom zero energy level (in red) in  Fig.~\ref{fig:uniform} is shifted by $\sqrt{2}t_1$ in the top spectrum.

These arguments apply to an infinite chain as well as to chains with periodic boundary conditions. In the case of a finite chain with open boundary conditions with an arbitrary number sites, impurity-like potentials may be generated at the edge sites of the sublattices when squaring the Hamiltonian (the local potentials are not uniform). However, for particular system sizes, the same reasoning can be applied. This will be discussed below, after we address the spectra of Sine-Cosine chains with periodic boundary conditions (PBC) in the next subsection.

\textbf{\emph{Application to bulk Hamiltonians}: }
In this subsection, we show that, when the unit cell of PBC Sine-Cosine chains has $2^{n}$ sites,  for certain choices of the ${\theta_j}$, the squaring process can be applied  $n$ times and, at each step, one of the  Hamiltonian blocks is again a Sine-Cosine chain (and this is why we call it a Matryoshka sequence), if an energy shift is applied and if the energy unit is renormalized. We label these  \textit{$n$-times squarable} Sine-Cosine chains, SSC$(n)$ [they are a  subset of the SC$(2^{n-1})$ chains].
Furthermore, the sequence of energy shifts and energy unit renormalizations determine the energy gaps in the respective spectrum.

A bulk Hamiltonian $H({k})$ is  the Hamiltonian of the unit cell closed onto itself with a twisted boundary (reflecting the $e^{ik}$ phases in the hopping terms  connecting one unit cell to the next), see Fig.~\ref{fig:sinecosine}(b). If the real space Hamiltonian is bipartite and the unit cell has more than one site, then $H(k)$ is also bipartite and the squaring process will generate a block diagonal matrix. 

Our Matryoshka sequence of Sine-Cosine chains is constructed starting from the last Hamiltonian in the squaring process which is that of a uniform chain with a single site in the unit cell and applying successively a square root operation (see Fig.~\ref{fig:uniform}). At each step of the square root process, we  obtain a Hamiltonian with a new chiral symmetry as illustrated in Fig.~\ref{fig:uniform} \cite{Asboth2016}.
The first  iteration deviates  from the general expressions for the following ones since the uniform chain has a  single-site  unit cell and therefore the respective bulk Hamiltonian cannot be  written in the sine-cosine form described in the beginning of this section. Therefore we describe this  first iteration before presenting the general expressions. 

\begin{enumerate}[leftmargin=*]
\item \textit{From the SSC$(0)$  to the SSC$(1)$ chain}. The Sine-Cosine chain SSC$(1)$  (corresponding to the SSH model) has hopping terms $\{\sin \theta, \cos \theta\}$ and when squared, generates a set of two equal bands with energy relation $\varepsilon(k)=1+\sin (2\theta) \cos k$ that corresponds to the spectrum of the  uniform tight-binding  SSC$(0)$ chain with an energy shift equal to one and a hopping parameter $t^{(0)} =\sin(2\theta) /\sqrt{2}$ (that determines the bandwidth). So the SSC$(0)$ chain has as band limits $\pm \sqrt{2} t^{(0)}$ and the chiral  level $\varepsilon_{SSC(0)}^{(0)}$ (folding level under the squaring operation) is zero. These values also determine the band structure of the SSC$(1)$ chain: the band limits are $\pm\sqrt{1\pm \sqrt{2} t^{(0)}}$ and a new chiral symmetry is present with chiral level  $\varepsilon_{SSC(1)}^{(1)}=0$.  The chiral level of the SSC$(0)$ chain is present at the SSC$(1)$ spectrum at the energies $\varepsilon_{SSC(0)}^{(1)}=\pm 1$. Note that the notation  $\varepsilon^{(n)}$ means a level in the spectrum of the SSC$(n)$ chain.

If we introduce a global factor $t^{(1)}$ in the hopping constants of the  SSC$(1)$ chain so that the hopping parameters become $\{t^{(1)} \sin \theta,t^{(1)} \cos \theta\}$, then the band limits become  $\pm t^{(1)}\sqrt{1\pm \sqrt{2} t^{(0)}}$ and $\varepsilon_{SSC(0)}^{(1)}=\pm t^{(1)}$.
Note that the uniform chain band energy shift and its bandwidth (for any choice of energy unit) determine the hopping parameters of the  SSC$(1)$ chain  and the same will occur if we repeat the square root operation [applying it  to the  SSC$(1)$ chain, then to  SSC$(2)$ and so on].
\begin{figure}[tb]
	\centering
	\includegraphics[width=0.90\linewidth]{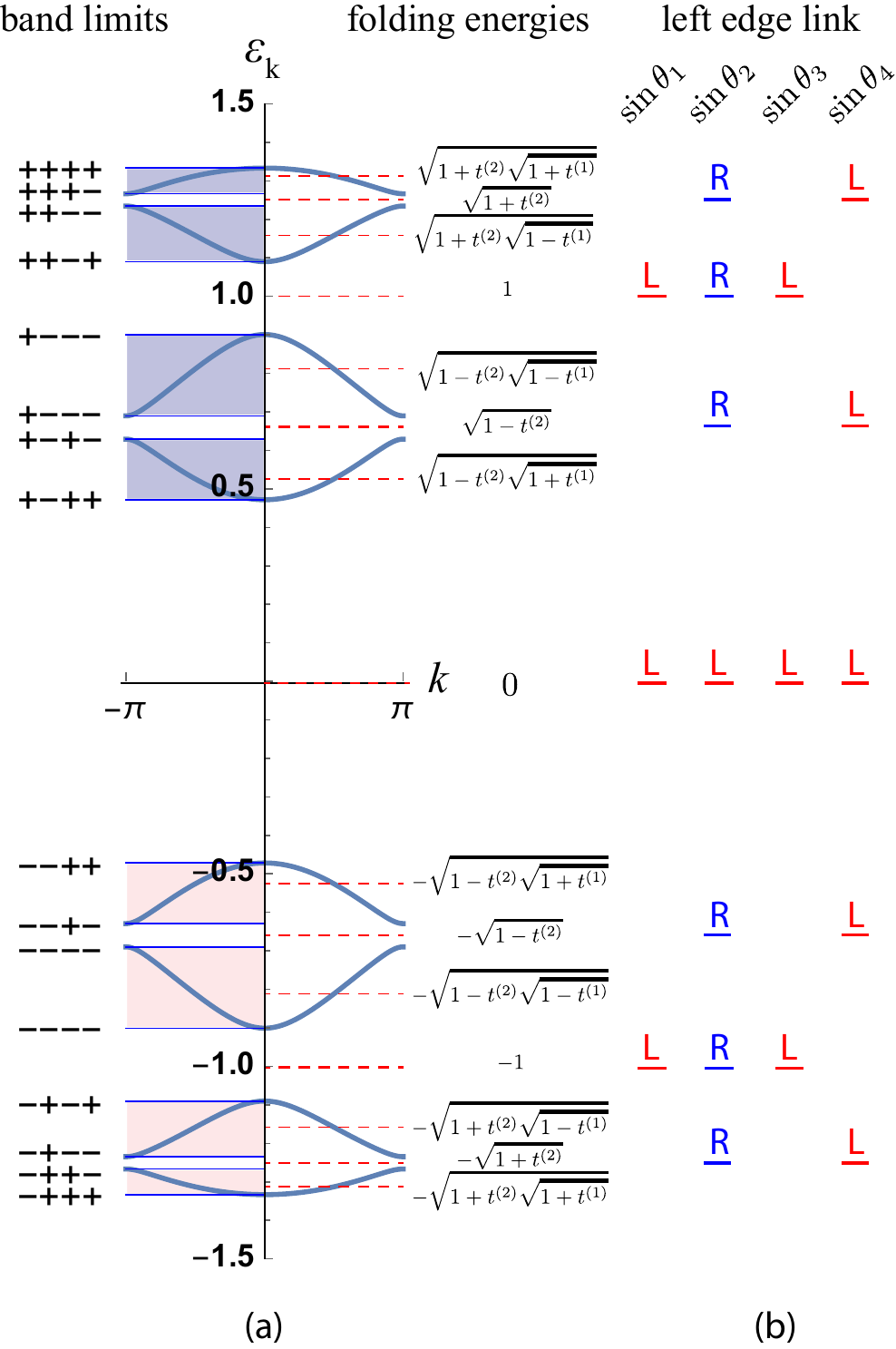}
	\caption{(a) Band structure of the SSC$(3)$ chain with $t^{(0)}=\sin (0.4 \pi)/ \sqrt{2}$, $t^{(1)}=0.9/ \sqrt{2}$, and  $t^{(2)}=0.8/ \sqrt{2}$ that generate unit cell hopping constants  $\{\sin \theta_1,\cos \theta_1,$ $\sin \theta_2,\cos \theta_2,$ $\sin \theta_3,$ $\cos \theta_3,$ $\sin \theta_4,\cos \theta_4\}\approx\{$0.542, 0.840, 0.309, 0.951, 0.485, 0.875, 0.375, 0.927$\}$. The folding levels (that intersect  energy curves at $\pm \pi/2$) are shown as well as the band limits $ \varepsilon_{\pm\pm \pm\pm}=$  $\pm  \sqrt{1\pm  t^{(2)} \sqrt{1\pm t^{(1)}\sqrt{1 \pm \sqrt{2}t^{(0}}}}$ (only the signs are indicated). (b)  Right (R) and left (L) edge levels of a SSC$(3)$  chain with OBC and $N =2^{n}p -1$ sites, with $n=3$ and  integer $p>1$, for all the possible choices of the leftmost hopping term that allow the squaring into  SSC$(j)$  chains.  }
	\label{fig:SC3}
\end{figure}

\item \textit{From the  SSC$(n-1)$ chain to the SSC$(n)$ chain}. 
When squaring the SSC$(n)$ Hamiltonian, the condition that it generates a block corresponding to a SSC$(n-1)$ chain is written as 
\begin{eqnarray}
	t^{(n-1)} \sin \theta_{j}^{(n-1)}&=& \cos \theta_{2j-1}^{(n)}  \sin \theta_{2j}^{(n)} 		\label{eq:1}\\
		t^{(n-1)} \cos \theta_{j}^{(n-1)}&=& \cos \theta_{2j}^{(n)}  \sin \theta_{2j+1}^{(n)} 
		\label{eq:2}
\end{eqnarray}
for $j=1, \cdots,$$2^{n-2}$ with $2^{n-1}+1$ $\equiv 1$. This implies that the global hopping factor in the  SSC$(n-1)$ chain is given by
\begin{equation}
	t^{(n-1)}= \sqrt{( \cos \theta_{2j-1}^{(n)}  \sin \theta_{2j}^{(n)})^2+(\cos \theta_{2j}^{(n)}  \sin \theta_{2j+1}^{(n)} )^2}
\end{equation}
for any value of  $j$.  These equations determine (almost uniquely, as we will explain further in the paper) the set $\{\theta_{j}^{(n)} \}$ of  the SSC$(n)$ if $t^{(n-1)}$ and $\{\theta_{j}^{(n-1)} \}$ are known.

Similarly to what was explained in the case SSC$(0)\rightarrow SSC(1)$, any level $\varepsilon^{(n-1)}$ in the SSC$(n-1)$ spectrum becomes a pair of levels ,  $\pm\sqrt{1+t^{(n-1)}\varepsilon^{(n-1)}}$, in the SSC$(n-1)$ spectrum. It is simple to conclude that the band structure of the SSC$(n)$ is characterized  by the following sequence of energy values that give the top and bottom energies of each band, 
\begin{multline}
\varepsilon_{\underbrace{\pm\pm \pm \cdots \pm\pm}_{n}}= \\
	\pm \sqrt{1\pm t^{(n-1)}\sqrt{1\pm t^{(n-2)}\sqrt{\ddots \sqrt{1\pm t^{(1)}\sqrt{1 \pm \sqrt{2}t^{(0)}}}}}},
	\label{eq:bandlim}
\end{multline}
where all the possible combinations of signs must be considered.
The folding levels associated with the chiral symmetries that appear at each step of the squaring process are, in the SSC$(n)$ spectrum, given by the ordered sequence of the values (all the possible combinations of signs must be considered)
\begin{align*}	
&\pm 1,\\
&\pm \sqrt{1\pm t^{(n-1)}},\\
&\pm \sqrt{1\pm t^{(n-1)}\sqrt{1\pm t^{(n-2)}}},\\
&\vdots,\\
&	\pm \sqrt{1\pm t^{(n-1)}\sqrt{1\pm t^{(n-2)}\sqrt{\ddots \sqrt{1\pm t^{(1)}}}}}.
\end{align*}

\end{enumerate}
To summarize, the set $\{\theta_{j}^{(n)} \}$ in the  SSC$(n)$ Hamiltonian is  determined  by the sequence of hopping factors $t^{(j)}$, $j=1,\cdots , n-1$, which are the energy units for each step of the construction of the  SSC$(n)$  chain, starting from the uniform chain. Obviously, all the bandwidths and gaps in the spectrum are also determined by this sequence. Note that we assumed that all hopping parameters are positive and this places all angles in the first quadrant. A gauge transformation can change the sign of the hopping terms maintaining the spectrum. Even with this condition,  the set $\{\theta_{j}^{(n)} \}$ in the  SSC$(n)$  is not unique given the sequence of hopping factors $t^{(j)}$, $j=1,\cdots , n-1$, because there are still the two possible choices for the  SSC$(j)$ sublattice at each step of the construction of the  SSC$(n)$  Hamiltonian (but the two choices generate the same spectrum).

The values of the angles in the caption of Fig.~\ref{fig:SC3} were derived using this procedure, that is, given the hopping parameters $t^{(0)}=\sin (0.4 \pi)/ \sqrt{2}$, $t^{(1)}=0.9/ \sqrt{2}$, and  $t^{(2)}=0.8/ \sqrt{2}$, we apply the procedure above three times starting from the uniform chain, generating at the first step the SSC$(1)$ chain (following the 1st step of the procedure), in the second the SSC(2) and so on (following the 2nd step of the procedure).

The first step is trivial since we defined the first hopping parameter as $t^{(0)}=\sin (0.4 \pi)/ \sqrt{2}$ and immediately we conclude that the SSC(1) Hamiltonian has hopping terms $\{\sin 0.2 \pi, \cos 0.2 \pi \}$. Note that in this case, when squaring the SSC$(1)$ Hamiltonian, the diagonal blocks are equal and therefore both blocks describe the same uniform chain. So, starting from the uniform chain, we have  two possible choices of sublattice corresponding to the initial uniform chain that imply the same SSC$(1)$ Hamiltonian except for a one-site shift of the hopping terms. Recall that we only consider angles in the first quadrant so that the hopping terms are positive. If we allowed for negative (or complex) hopping terms, other solutions would be obtained corresponding to gauge transformations of the Hamiltonian with positive hopping terms.

The SSC$(j)$ Hamiltonian is obtained from the SSC$(j-1)$ one, with $j>1$, solving numerically Eqs.~\ref{eq:1} and \ref{eq:2} in order to find the angles $\{\theta^{(j)}_i\}$, given the hopping parameters $t^{(j)}$, $j=0, …, n-1$. In the construction of the SSC$(3)$ bulk Hamiltonian, the spectrum of which is shown in Fig.\ref{fig:SC3}, two different solutions (but that generate the same SSC$(j-1)$ diagonal block when squared and the same spectrum) were found at each of the steps SSC$(1)\rightarrow$SSC$(2)$ and SSC$(2)\rightarrow$SSC$(3)$. In the first of these steps, we chose the odd sublattice as the one supporting the lower order SSC$(j-1)$ Hamiltonian and in the second, we chose the even one. The hopping constants  indicated in Fig.~\ref{fig:SC3} reflect a choice of one of those two solutions at each of these steps. The existence of two solutions is easy to justify in the SSC$(1)\rightarrow$SSC$(2)$ step, noting that the SSC$(2)$ bulk Hamiltonian with the hopping sequence $\{ \sin \theta_1,  \cos \theta_1,  \sin \theta_2,  \cos \theta_2 e^{-ik}\}$ generates the same SSC$(1)$ block as the one with the hopping sequence $\{ \cos \theta_2,  \sin \theta_2,  \cos \theta_1,  \sin \theta_1 e^{-ik}\}$.

\begin{figure}[tb]
	\centering
	\includegraphics[width=0.90\linewidth]{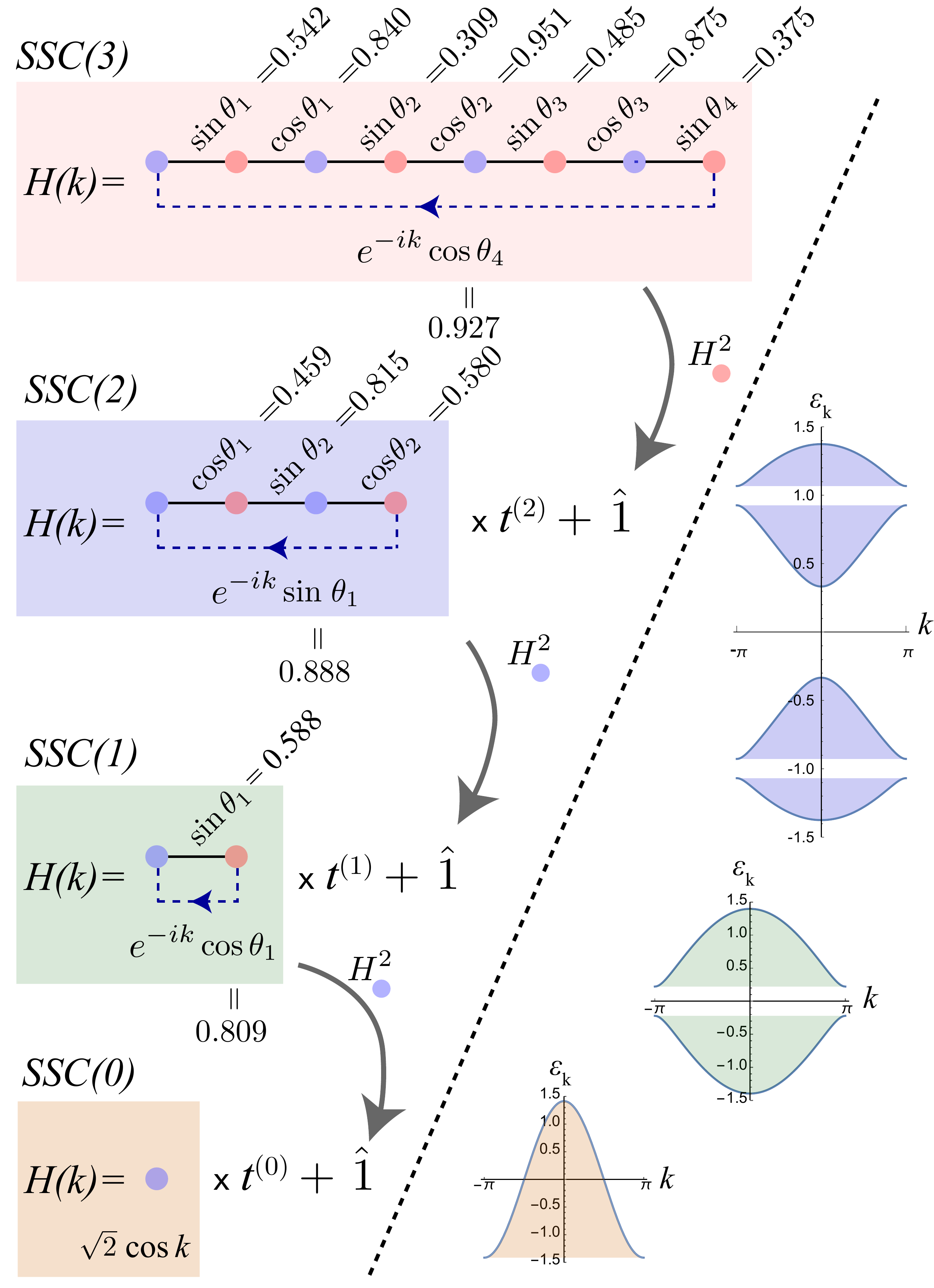}
	\caption{Matryoshka of bulk SSC$(j)$ Hamiltonians associated with the construction of the SSC$(3)$ Hamiltonian  with spectrum shown in Fig.~\ref{fig:SC3}.  The sublattice corresponding to the SSC$(j-1)$ is indicated by the color of the index site in the arrow $H^2$. 
	The  band structure of the  SSC$(j)$ chains is shown on the bottom right corner [the corresponding SSC$(j)$ chain is indicated by the filling color]. }
	\label{fig:SC3contruct}
\end{figure}

The numerical solution can be checked by: (i) squaring the SSC$(3)$ Hamiltonian; (ii) selecting the diagonal block with unitary local energies; (iii) removing the diagonal matrix elements of this block; (iv) multiplying the block by a constant (the inverse of $t^{(j-1)}$) that reduces this block to the SSC$(j-1)$ Hamiltonian form. Repeating this procedure three times, one obtains the bulk SSC$(j)$ Hamiltonians associated with the construction of the SSC$(3)$ Hamiltonian  with spectrum shown in Fig.~\ref{fig:SC3}. This Matryoshka sequence is shown in Fig.~\ref{fig:SC3contruct}, as well as the respective band structure of the  SSC$(j)$ chains.

Note that the  SSC$(n)$ chain can be viewed as a $2^n$-root of a uniform tight-binding chain, but a generalized one due to  degrees of freedom associated with global hopping factors $t^{(n)}$.

\textbf{\emph{Finite systems with open boundaries}: }
In this section, we show how to generate edge states in any of the gaps in the band structure of the SSC$(n)$ chain which will be protected by the chiral symmetry of a  particular step of the construction of the  SSC$(n)$  Hamiltonian. Note that the edge states are replicated in the correspondent gaps in the unfolding process. For example, the gaps in Fig.~\ref{fig:SC3} from top to bottom are  SSC$(1)$,  SSC$(2)$,  SSC$(1)$,  SSC$(3)$, SSC$(1)$,  SSC$(2)$,  SSC$(1)$  gaps and an edge state associated with the  SSC$(1)$  chain will be present in all  SSC$(1)$  gaps.

Let us first explain the appearance of edge states in the usual SSH(2) chain [equivalent to the SC$(1)\equiv SSC(1)$ chain].
Edge states appear in an open boundary SSH(2) chain when a weak link is present at the boundaries. Our definition of weak link is a hopping term in the unit cell that can be adiabatically increased from zero (with all the other hopping terms in the unit cell finite, constant and larger) without  closing the central gap.  If one of the sublattices of the bipartite chain has one more site, an edge state is always present (it changes from a right edge state to a left edge state at the topological transition, reflecting the fact that there is always one weak link at one of the boundaries) and has finite support only in this sublattice. Furthermore, its energy is exactly zero, a value protected by the chiral symmetry. If both sublattices have the same number of sites and we can split the chain in two halves, each of them with a weak link at the boundary, two edge states will be present with nearly zero energy which are protected by the chiral symmetry and the band gap. 

In order to generate edge states in a chosen gap of the SSC$(n)$ chain, one chooses the boundaries  in such a way that the squaring process will generate the SSC$(j)$ chain  that has that gap as the central one with a weak link at its boundaries. Also, in order to guarantee that one of the blocks at each squaring step is that of a bipartite chain (apart from an energy shift), we impose that the number of sites of the SSC$(n)$ chain is $N =2^{n}p -1$, with integer $p>1 $, so that the inner sublattice is the one corresponding to the bipartite  OBC SSC$(j)$ chain in all steps (the number of sites at each step is of the form $N =2^{j}p -1$ ). That way, all sites of the  OBC SSC$(j)$ chain will have the same local potential (equal to one). In the case of the SSC$(n)$ chain, there are $2^{n-1}$ possible choices of chain terminations given a system size $N =2^{n}p -1$. In Fig.~\ref{fig:SC3}(b), we show the numerically determined edge state levels in the case of a OBC SSC(3) chain with $N =2^{n}p -1$ sites for all the possible choices of the leftmost hopping term ($\sin \theta_1$, $\sin \theta_2$, $\sin \theta_3$, $\sin \theta_4$). The observed edge state levels agree with the previous argument, that is, when the left edge link is, for example,   $\sin \theta_3$,  squaring once the Hamiltonian, one obtains a block diagonal matrix where one of the blocks corresponds to the SSC$(2)$ chain with a single weak link at the left edge and this justifies the existence of a left edge state in the SSC$(2)$ gaps of the OBC SSC$(3)$ chain.

Interestingly, for these system sizes, the squaring method can be extended until we reach  a single level. This implies that  the spectrum of the OBC SSC$(n)$ chain is the combination of:
(i) the spectrum obtained from a single level with zero energy, applying successively energy shifts, energy unit renormalizations and square roots to each level [each level $\varepsilon^{(j-1)}$ generates $\pm\sqrt{1+t^{(j-1)}\varepsilon^{(j-1)}}$ levels in the SSC$(j)$ spectrum]; (ii) a spectrum that has levels only at the folding energies (due to the extra site in the other sublattice relatively to the SSC$(j)$ sublattice).

The presence of edge states in the central band at each step of the construction can be confirmed adding the Zak's phase of the positive bands leading as usual to $\pi$ in the non-trivial topological phase.

\begin{figure}[tb]
	\centering
	\includegraphics[width=0.90\linewidth]{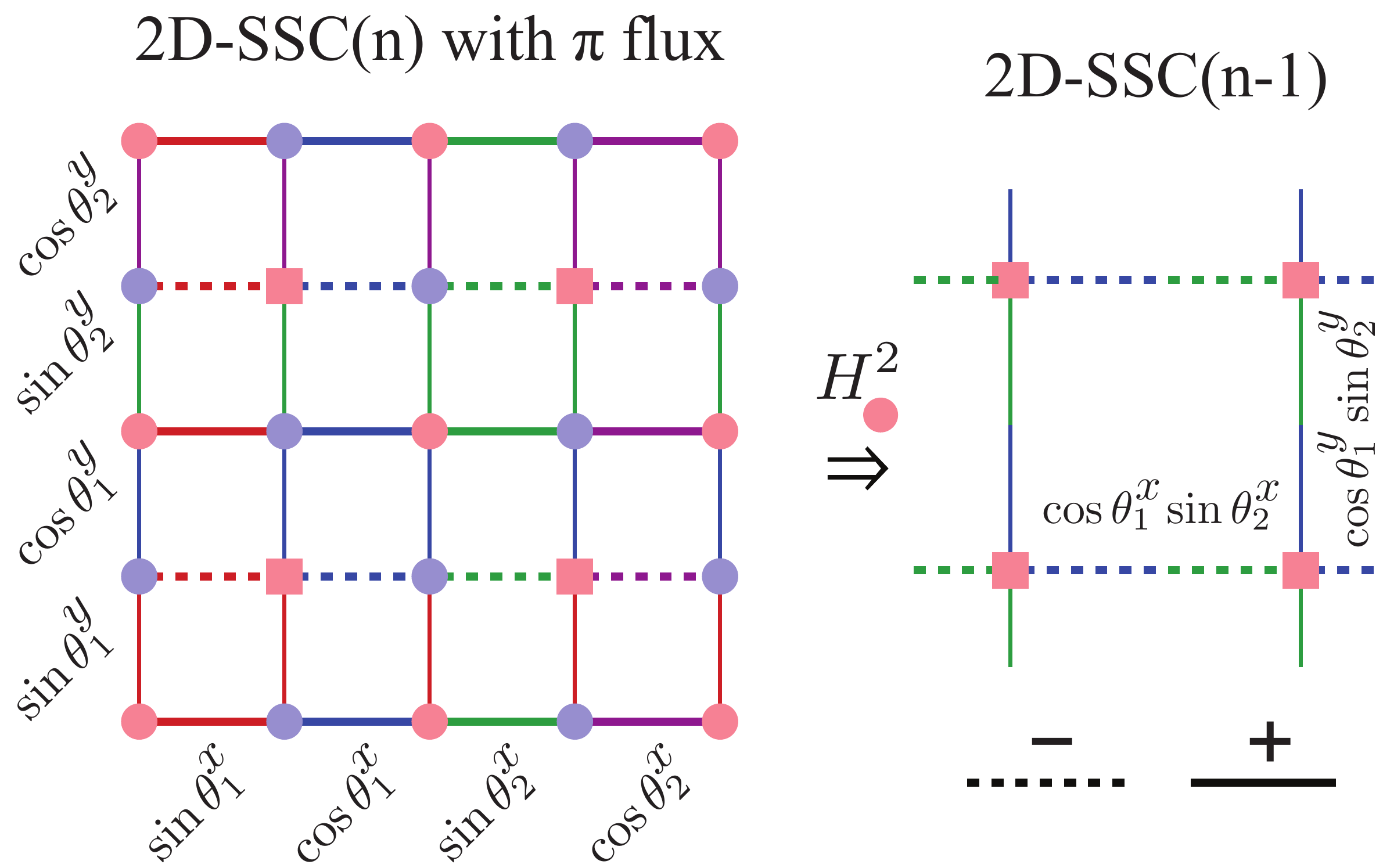}
	\caption{A square-root Hamiltonian of a 2D SSC$(n-1)$ Hamiltonian is possible if $\pi$-flux per plaquette is introduced in the  2D SSC$(n)$ lattice. This flux implies the existence of 4 blocks (four decoupled sublattices) in the squared Hamiltonian, one of them being the 2D SSC$(n-1)$ one (sublattice with squared sites).  }
	\label{fig:2D}
\end{figure}
\textbf{\emph{Extension to 2D}: }
The 2D version (for higher dimensions the reasoning is similar) of the SSC$(n)$  chain can be constructed in the same way as the 2D SSH model is constructed from the SSH chain, that is, the hopping terms in the $x$ direction are those of a $x$-SSC$(n)$ chain and the same for  hopping terms in the $y$ direction [the $y$-SSC$(n)$ hopping terms can be different from the $x$-SSC$(n)$ ones, as in the asymmetric 2D SSH model \cite{Pelegri2019}]. This 2D model will have a band structure that can be characterized by the band limits at the inversion invariant momenta and these limits will be the sum of two terms of the form of Eq.~	\ref{eq:bandlim}, $\varepsilon^x_{\pm\pm \pm \cdots \pm\pm}+\varepsilon^y_{\pm\pm \pm \cdots \pm\pm}$.

One may be tempted to try to construct the 2D-SSC$(n)$ model following the method given for the SSC$(n)$ chain so that, when squaring the Hamiltonian, 2D-SSC$(j)$ blocks are generated. Despite the fact that the lattice is bipartite, one faces one difficulty: the dimension of the 2D-SSC$(n-1)$ model is one fourth of that of the 2D-SSC$(n)$ model. When squaring the Hamiltonian, the bipartite property guarantees the appearance of two diagonal blocks, each one corresponding  to different sublattice (in Fig.~\ref{fig:2D}, the two sublattices have different colors). An extra factor is required in order for one of these blocks to become a diagonal sum of two smaller blocks, reflecting the division of the sublattice in two other sublattices (with  pink circular sites and pink squared sites in Fig.~\ref{fig:2D}, respectively). So we are only able to find a single square-root of a 2D SSC$(n-1)$ model, and that is the 2D SSC$(n)$ model with $\pi$-flux per plaquette, with flux introduced by multiplying the $x$-hopping terms by $(-1)$ at every other rung.
This $\pi$ flux generates destructive interference in the squared hopping terms between  pink circular sites and pink square sites in Fig.~\ref{fig:2D} and therefore one may interpret it as an additional ``bipartite'' property.

This extended 2D version of the SSC$(n)$ chain can be considered a generalization of the Benalcazar-Bernevig-Hughes model \cite{Benalcazar2017, Benalcazar2017a}. This latter model is a higher-order bipartite topological insulator with corner states at the central gap. 
In our 2D-SSC$(n-1)$ model of Fig.~\ref{fig:2D}, if at a corner site we have weak links in the $x$ and $y$ directions, a zero-energy corner state will be present. However, no central bulk band gap will be present due to the $C_{4v}$ symmetry when the hopping parameters in $x$ and $y$ directions are the same \cite{Cerjan2020,Benalcazar2020} and one would expect this zero-energy corner state not to be protected. Recently, it has been shown that combination of the $C_{4v}$ and chiral symmetries protect these zero-energy corner states from hybridizing with the bulk states \cite{Cerjan2020,Benalcazar2020}.
The existence of the zero-energy corner state in the spectrum of the 2D-SSC$(n-1)$ model of Fig.~\ref{fig:2D}  implies the presence of finite energy corner states in the continuum of the non-central energy bands of the 2D-SSC$(n)$ model of Fig.~\ref{fig:2D} (the respective energies indicating where possible gaps could appear, for example, introducing a magnetic flux in the 2D-SSC$(n-1)$ model). 
The symmetry protection of the zero-energy corner states of the 2D-SSC$(n-1)$ model implies the same protection of the finite energy corner states of the 2D-SSC(n) model.

\textbf{\emph{Conclusion}: }
Square-root topological insulators have attracted attention due to the presence of finite energy topological edge states in non-central gaps of the chiral spectrum that cannot be characterized using the usual topological invariants. In this paper, we extend the concept of $\sqrt{TI}$ by introducing a particular 1D Hamiltonian [that we label $n$-times squarable Sine-Cosine model, SSC$(n)$]  of the family of the SSH($2^n$) chains that can be squared multiple times generating at each step a self-similar Hamiltonian (with a smaller unit cell) in what we call a Matryoshka sequence, each of them with its own chiral symmetry. Edge states at any gap of the original chain are protected by one of these chiral symmetries. 
Given this model, one may extend perturbatively the respective topological characterization to other (not sine-cosine) models that are generated from a sine-cosine chain modifying the hopping parameters, as long as these modifications are small compared with the energy gaps. Zero and finite energy edge states will be robust under such modifications.
Furthermore, 2D weak topological insulators are 1D topological insulators for a fixed transverse momentum and also here, it is possible to interpret the existence of finite energy edge states using this perturbed sine-cosine model argument.

These models are determined by the sequence of energy unit renormalizations in the squaring process and their spectrum has a very simple form in terms of these parameters. This fine tuning of their band structure as well as the control over the presence or absence of edge states in any of the spectrum gaps makes these models very appealing in the context of artificial lattices such as photonic \cite{Kremer2020,Baboux2016,Mukherjee2018,Mukherjee2020,Xia2020,Jorg2020}, optical lattices \cite{Aidelsburger2013,Jotzu2014,Tale2015}, topoelectrical circuits \cite{Albert2015,Olekhno2020} or acoustical lattices \cite{Chen2021,Li2018}, where the effective hopping terms can be adjusted in order to reproduce the necessary set of angles $\{\theta_j\} $. In particular, topoelectrical circuits may be the best choice of artificial lattice for the simulation of the sine-cosine chains studied in our paper since in these systems the hopping and on-site potential parameters depend only on the capacitance and inductance values of the capacitors and inductors which can be controlled very precisely. In the case of ultra-cold quantum gases in optical lattices, tunneling terms can also be controlled modifying the potential wells. 

In this paper, we did not discuss the possibility of disorder on the onsite energies or on the hopping matrix elements. However, from perturbation theory one can state that the edge states of the SSC$(n)$ chain will be robust against the introduction of disorder as long as the fluctuations of the onsite energies or of the hopping matrix elements are small compared with all band gaps.

\textbf{\emph{Acknowledgments}: }
This work was developed within the scope of the Portuguese Institute for Nanostructures, Nanomodelling and Nanofabrication (i3N) projects UIDB/50025/2020 and UIDP/50025/2020. RGD and AMM acknowledge funding from FCT - Portuguese Foundation for Science and Technology through the project PTDC/FIS-MAC/29291/2017. AMM acknowledges financial support from the FCT through the work contract CDL-CTTRI-147-ARH/2018.

\bibliography{bibliografia2}

\end{document}